\documentclass{elsart4}

\usepackage{graphics}

\usepackage{amssymb}

\usepackage[english,francais]{babel}

\newtheorem{e-proposition}[theorem]{Proposition}

\newtheorem{e-definition}[theorem]{Definition\rm}

\renewcommand{\theequation}{\arabic{equation}}
\def\og{\leavevmode\raise.3ex\hbox{$\scriptscriptstyle\langle\!\langle$~}}
\def\fg{\leavevmode\raise.3ex\hbox{~$\!\scriptscriptstyle\,\rangle\!\rangle$}}

\def\spose#1{\hbox to 0pt{#1\hss}}
\def\gsim{\mathrel{\spose{\lower 3pt\hbox{$\mathchar"218$}}
          \raise 2.0pt\hbox{$\mathchar"13E$}}}
\def\lsim{\mathrel{\spose{\lower 3pt\hbox{$\mathchar"218$}}
          \raise 2.0pt\hbox{$\mathchar"13C$}}}

\begin{document}
\centerline{Title of the dossier/Titre du dossier}
\begin{frontmatter}


\selectlanguage{english}
\title{Afterglows after Swift}


\selectlanguage{english}
\author[authorlabel1]{O. Godet},
\ead{olivier.godet@cesr.fr}
\author[authorlabel2]{R. Mochkovitch}
\ead{mochko@iap.fr}

\address[authorlabel1]{Universite Toulouse III, Institut de Recherche 
d'Astrophysique \& de Planetologie (IRAP)\\ 9 Av. du Colonel Roche,
31028 Toulouse Cedex 4, France}

\address[authorlabel2]{Institut d'Astrophysique de Paris, UMR 7095,
CNRS -- Universit\'e Pierre et Marie Curie\\
98bis Bd Arago, 75014 Paris, France}

\begin{abstract}
Since their discovery by the Beppo-SAX satellite in 1997, gamma-ray burst
afterglows have attracted an ever-growing interest.  They have
allowed redshift measurements that have confirmed that gamma-ray bursts are located at
cosmological distances. Their study covers a huge range both in time (from one
minute to several months after the trigger) and energy (from the GeV to radio
domains).  The purpose of this review is first to give a short historical
account of afterglow research and describe the main observational results
with a special attention to the early afterglow revealed by {\it Swift}. We
then present the standard afterglow model as it has been developed in the
pre-{\it Swift} era and show how it is challenged by the recent {\it Swift}
and {\it Fermi} results. We finally discuss different options (within the
standard framework or implying a change of paradigm) that have been proposed
to solve the current problems.

{\it To cite this article: O. Godet, R. Mochkovitch, C. R. Physique 12 (2011).}

\vskip 0.5\baselineskip

\selectlanguage{francais}
\noindent{\bf R\'esum\'e}
\vskip 0.5\baselineskip
\noindent
{\bf Les r\'emanences apr\`es Swift.}  Depuis leur d\'ecouverte par le
satelite Beppo-SAX en 1997, les r\'emanences des sursauts gamma ont suscit\'e
un int\'er\^et sans cesse croissant. Elles ont permis les mesures de redshifts
qui ont confirm\'e que les sursauts sont situ\'es \`a distance
cosmologique. Leur \'etude couvre un vaste domaine aussi bien temporel (de une
minute \`a plusieurs mois apr\`es le sursaut) que spectral (depuis les GeV
jusqu'\`a la radio).  Le but de cette revue est d'abord de faire un court
rappel historique de la recherche sur les r\'emanences puis de d\'ecrire les
principaux r\'esultats observationnels avec une attention particuli\`ere pour
la r\'emanence pr\'ecoce, r\'ev\'el\'ee par {\it Swift}. Nous pr\'esentons
ensuite le mod\`ele standard tel qu'il fut d\'evelopp\'e durant l'\`ere
pr\'e-{\it Swift} et montrons comment il est mis en d\'efaut par les
r\'esultats r\'ecents de {\it Swift} et {\it Fermi}. Nous discutons enfin les
propositions (dans le cadre standard ou impliquant au contraire
un changement de paradigme) qui ont \'et\'e avanc\'ees pour r\'esoudre les
diff\'erents probl\`emes.

{\it Pour citer cet article~: O.Godet, R. Mochkovitch, C. R. Physique 12 (2011).}

\keyword{Gamma-Ray Bursts; Afterglow; Hydrodynamics; Radiative Processes;
Multiwavelength Observations} \vskip 0.5\baselineskip
\noindent{\small{\it Mots-cl\'es~:} Sursauts Gamma; R\'emanence; Hydrodynamique;
Processus Radiatifs; Observations Multi-longueurs d'onde}}
\end{abstract}
\end{frontmatter}

\selectlanguage{english}
\section{Introduction}

Even before the discovery of the first afterglow by Beppo-SAX it was suspected
that the prompt phase of gamma-ray bursts (hereafter GRBs) should be followed by some long
lasting emissions at longer wavelengths.  By analogy with supernova remnants it
seemed natural to believe that a fraction of the energy dissipated when the
relativistic ejecta is decelerated by the circumburst medium would be
radiated, giving rise to an ``afterglow''.

The detection of the first X-ray afterglow associated with 
GRB\,970228 on February
 28th, 1997 marked a milestone in the GRB study \cite{Costa97}. The 3
 arcminute X-ray error box was searched with the William Herschel
 and Isaac Newton telescopes at La Palma and a weak optical counterpart was
 discovered. On March 1$^{\rm st}$ it had a visual magnitude $V = 21.3$ and
 declined below 24 after a week. A faint galaxy ($R = 24$) was found within
 0.2 arcsecond from the counterpart. It was the likely host of the burst and
 this, for the first time, provided a direct evidence that GRBs were located
 at very large (cosmological) distances. 
Three months later another burst (GRB 970508)
was localized by Beppo-SAX and a spectrum of the afterglow taken by the Keck 
telescope showed redshifted lines
of silicon and iron at $z = 0.77$ and $z = 0.835$ (the largest redshift 
being that of the host galaxy and 
the smaller one corresponding to an intervening absorption system). This 
first redshift definitely confirmed 
the cosmological
distance scale. In December of the same year GRB 971214 reached $z = 3.42$.

From that moment, it became possible to identify long-GRB host
 galaxies leading in turn to constraints on GRB progenitors, to have access to
 physical quantities that enabled to intensively test theoretical models. This
 outstanding achievement was made possible thanks to the Italian-Dutch
 BeppoSAX (e.g. \cite{Scarsi93}) mission 
with the spacecraft being able to slew within a few hours of
an alert to bring the GRB trigger position in the field of view of
the narrow field X-ray instruments (NFIs) that could further refine the
error box.
This allowed, for the first time, ground
 facilities to observe quickly the GRB multi-wavelength afterglow. 
 HETE 2 \cite{Lamb}, launched in 2000, had similar capabilities and 
the two satellites collected a total of nearly 40 redshifts from 1997 to 2005.

 Another
 milestone in GRB study was marked by the launch of the {\it Swift} mission
 \cite{Gehrels04} in November 2004 that brings the BeppoSAX and HETE 2 observing
 strategy to the best refinement 
by associating a slew capability of a few tens
of seconds with a platform carrying a wide-field coded-mask telescope
providing the trigger and two very sensitive NFIs in X-rays and optical.
This led {\it Swift}:
 i) to fully characterise the emission, in particular in X-rays, 
from a temporal interval (between a few minutes and a few
hours from the prompt emission) that was previously not accessible, 
the so-called {\it early afterglow}; ii) to discover the multi-wavelength afterglows of the
 prior-elusive short GRBs, which led in turn to better understand their
 nature; iii) to detect several high-redshift GRBs of which 3 events with $z =
 6.3,~6.7$ and 8.2.

GRB afterglows convey broadband information on the physics of the GRB
phenomenon itself, but also on the GRB host galaxies and the GRB environment
and hence on the GRB progenitors, and as cosmological probes they allow to
study the evolution of baryons with redshift or even possibly the
re-ionisation of the Universe for the furthest ones. Here, we discuss about
our current state-of-the-art understanding of the afterglow physics in the
{\it Swift} and {\it Fermi} era.  The use of GRBs as cosmological tools as
well as the nature of their progenitors is addressed respectively in \cite{PP11} \& \cite{BZ11} (this issue).  
We focus on the behaviour of the afterglow
in X-rays that was intensively characterised with {\it Swift} and the related
physics, and then we discuss how the observations in other wavelengths fit or
not in this framework.

The review is organised as follows: in Section 2, we describe the standard
afterglow model. In Section 3, we briefly present the {\it Swift} mission and
then we detail the afterglow X-ray canonical light curve revealed by {\it
Swift}. We discuss in some length its physical interpretation as well as what
observations in other wavelengths tell us. We discuss in Section 4 the
afterglow properties of different classes of GRBs. In Section 5, we give our
view on some of the outstanding current open issues that future missions and
theoretical developments will need to address.

\section{Standard model of the afterglow}
\label{model}

Following the discovery of the first afterglows, models were readily
constructed to interpret the new wealth of data.  Any model of the afterglow
consists in two parts: (i) an hydrodynamical description of the deceleration
of the relativistic burst ejecta by the circumstellar medium and (ii) the
identification of the radiative processes at work.  The deceleration leads to
the usual two-shock picture: a forward shock propagating in the external
medium and a reverse shock sweeping back into the ejecta, the two shocked
regions being separated by a contact discontinuity. The standard model
assumes that the afterglow results from the synchrotron emission of electrons
accelerated in the forward shock \cite{MR97}. In the basic version of the
model the hydrodynamics is described by the Blandford-McKee
\cite{BlandfordMcKee76} solution (a relativistic analogue to the classical
Sedov self-similar blast wave model). Behind the shock, the electrons are
accelerated to a non-thermal power law distribution and then emit synchrotron
radiation in the local magnetic field. The dynamics is fixed by the total
explosion kinetic energy $E_{\rm K}$ (in erg sr$^{-1}$ since the ejecta is
supposed to be beamed in two opposite jets of initial opening angle
$\theta_j$) and the external density (a uniform or wind-like
medium). The instantaneous synchrotron spectrum at a given shock location
depends on three physical quantities: the slope $p$ and minimum Lorentz factor
$\gamma_m$ of the initial distribution of electrons, and the magnetic field
strength $B$. The last two are estimated by assuming that respective fractions
$\epsilon_e$ and $\epsilon_B$ of the dissipated energy (in the local fluid
frame) are used to accelerate electrons and amplify the magnetic
field \cite{LP11}. Comparing the hydrodynamical expansion time to the synchrotron cooling
time then leads to define the Lorentz factor $\gamma_c$: electrons with
Lorentz factors $\gamma > \gamma_c$ can radiate efficiently while those with
$\gamma < \gamma_c$ mostly loose their energy by adiabatic expansion.  Hence,
depending on the relative order of $\gamma_m$ and $\gamma_c$ one distinguishes
two cooling regimes: fast cooling ($\gamma_m > \gamma_c$) and slow cooling
($\gamma_m < \gamma_c$). \cite{Sari98} obtained the expected spectra (for a
power law distribution of injected electrons) in the two cases. Considering
the synchrotron frequencies $\nu_m$ and $\nu_c$ respectively associated to
$\gamma_m$ and $\gamma_c$ ($\nu_{\rm syn}\propto B\gamma^2$) the elementary
spectra are made of three power law segments for $\nu$ $<$ min($\nu_m$,
$\nu_c$), min($\nu_m$, $\nu_c$) $<$ $\nu$ $<$ max($\nu_m$, $\nu_c$) and $\nu$
$>$ max($\nu_m$, $\nu_c$). A fourth segment appears at low frequency when the
possibility of self-absorption of the synchrotron radiation by the emitting
electrons is taken into account. Coupling this result to the Blandford-McKee
solution it is then possible to predict the afterglow light curve at a given frequency
($F_{\nu}\propto t^{-\alpha}~\nu^{-\beta}$)
which is also made of several power law segments (with 
the so-called closure relations
coupling the temporal and spectral indices $\alpha$ and $\beta$, see e.g. \cite{Zhang06}). 
This was first done considering
a forward shock propagating into a constant density interstellar medium 
(ISM)\cite{Sari98}. \cite{LiChevalier01} and \cite{PK00} then obtained similar solutions in
a stellar wind environment. During the pre-{\it Swift} era this
simple model appeared to give a fairly good account of the available
observations
(e.g. \cite{Panaitescu04a,Panaitescu04b,Panaitescu01}). Multiwavelength fits
of the data were used to constrain burst parameters such as the kinetic
energy, external density or $\epsilon_e$, $\epsilon_B$ and $p$.  Surprisingly,
a uniform external medium appeared to be preferred in most cases while the
progenitors of long bursts (the only ones with detailed afterglow observations
in the pre-{\it Swift} era) are supposed to be Wolf-Rayet stars, which are
normally thought to have strong winds. Another potential problem comes from
the relatively high efficiency of the prompt emission mechanism implied by
these results: the ratio $E_{\gamma}/E_{\rm K}$ between the energy measured in
gamma-rays and the kinetic energy obtained from the fits is often larger than
10\%. Such an efficiency is difficult to reach by internal shocks, the most
discussed mechanism for the prompt emission.

\section{The Swift era}
\subsection{The Swift mission}

The {\it Swift} mission \cite{Gehrels04} is a NASA, Italian and UK satellite
launched in November 2004 to study GRBs with several challenging goals,
amongst them: i) the detection of high-redshift GRBs; ii) the identification
and characterisation of the afterglows from short GRBs; iii) the sampling of
the afterglow emission in X-rays and optical from a few tens of seconds after
the trigger to several days, thanks to rapid slew capability of about 1-2
minutes.  To do so, {\it Swift} embarked three instruments: the Burst Alert
Telescope (BAT \cite{Barthelmy05}), a 15-150 keV and 2\,sr coded-mask camera
with a 5200 cm$^2$ geometrical area in charge of the detection and primary
localisation of GRBs with a typical accuracy of 3$^{\prime}$ in radius at a
90\% confidence level and two narrow-field instruments. The X-Ray Telescope
(XRT \cite{Burrows05}) and the UV/Optical Telescope (UVOT \cite{Roming05})
enable to refine the GRB error box after the spacecraft slew with typical
accuracy in radius of better than 3$^{''}$ and $< 1^{''}$ at a 90\% confidence
level, respectively. The XRT is a 0.3-10 keV Wolter-I telescope with a focal
length of 3.5 m, an effective area of 110 cm$^2$ at 1.5 keV and a field of
view of 23.6 $\times$ 23.6 arcminutes. The UVOT is a modified
Ritchey-Chr\'etien telescope with a field of view of 17 $\times$ 17 arc-minute
covering the 170-650 nm range.

\begin{figure*}\center{
\resizebox{.6\hsize}{!}{\includegraphics{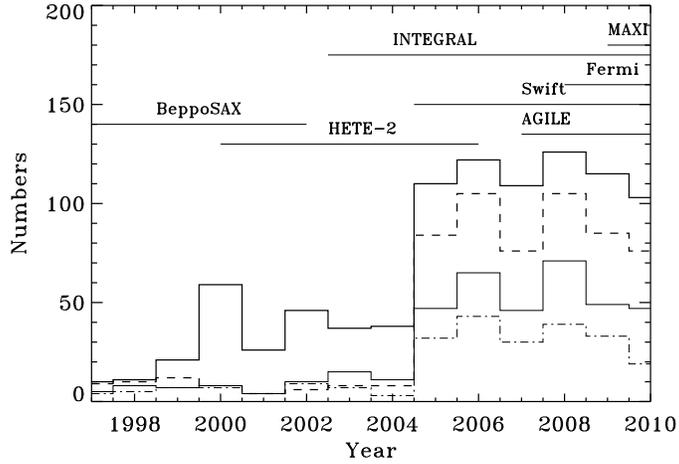}}}

\caption{Plot showing the number of GRBs (thick solid line) detected by
various gamma-ray instruments from 1997 to 2010, the number of GRBs with
detected X-ray (thick dashed line) and optical (thin solid line) afterglows,
the number of GRBs with measured spectroscopic redshift (thin dotted-dashed 
line). \label{fig_stat1}}
\end{figure*}

\begin{figure*}\center{
\resizebox{.6\hsize}{!}{\includegraphics{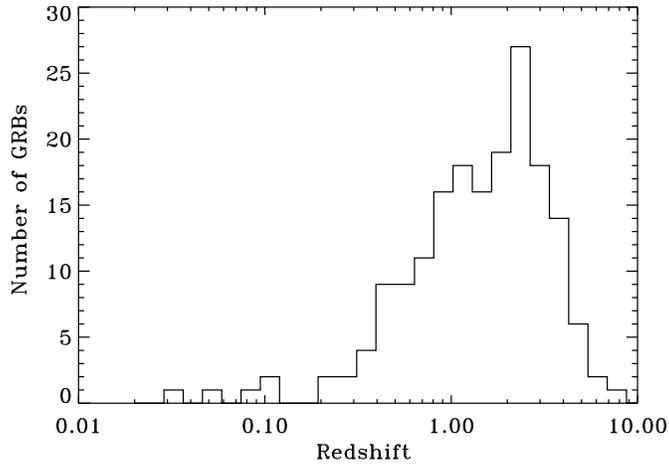}}}

\caption{Distribution of measured spectroscopic redshifts up to the end of
  2010.  \label{fig_stat2}}
\end{figure*}

Excluding the GRBs found by BATSE, 915 GRBs have been detected from 1997 to
2010. From this total, $\sim$ 65\%, $\sim$ 43\% and $\sim$ 8\% have been also
seen in X-rays, IR/Optical and radio, respectively (see Fig.~\ref{fig_stat1}).
From its launch to the end of the year 2010, {\it Swift} found 547 GRBs
($\sim$ 60\% of the 915 GRBs since 1997) of which more than $90\%$ were also
observed by the XRT and only $\sim 40\%$ by the UVOT. It is interesting to
note that even if ground follow-ups increase the fraction of GRBs detected in
IR/Optical by $\sim 15-20\%$, this fraction is still less than that observed
in X-rays. The deficit is not the result of a lack of sensitivity of the UVOT
and ground instruments, but more intrinsic to GRBs which appear for some of
them as dark in IR/optical (e.g. \cite{Rol05,Roming06} -
cf. Section~\ref{optical}). As seen in Fig. 1 and 
~\ref{fig_stat2}, not only {\it  Swift} brought a large sample of GRBs, it also increased by a factor of
about 3 the number of GRBs with a measured spectroscopic redshift. The average
redshift for {\it Swift} GRBs is around 2.0 (using spectrospic redshift
measurements from 2005 to the end of 2010) against 1.2 prior to {\it Swift}
(see e.g. \cite{Mangano06}). This results from the higher sensitivity of the
BAT that detects fainter GRBs when compared to previous GRB trigger
cameras. Another tremendous achievement of {\it Swift} is its capability to
follow the afterglow emission from a few tens of seconds (in some cases at a
time when the prompt gamma-ray ray emission is still ongoing) to several days
or couple of weeks and in some exceptional cases over more than a month as for
GRB 060729 (see Fig.~\ref{fig_example_lc}; \cite{Grupe07}).

\subsection{Swift X-ray observations}
\subsubsection{The canonical X-ray light curve}

\begin{figure*}\center{
\resizebox{.8\hsize}{!}{\includegraphics{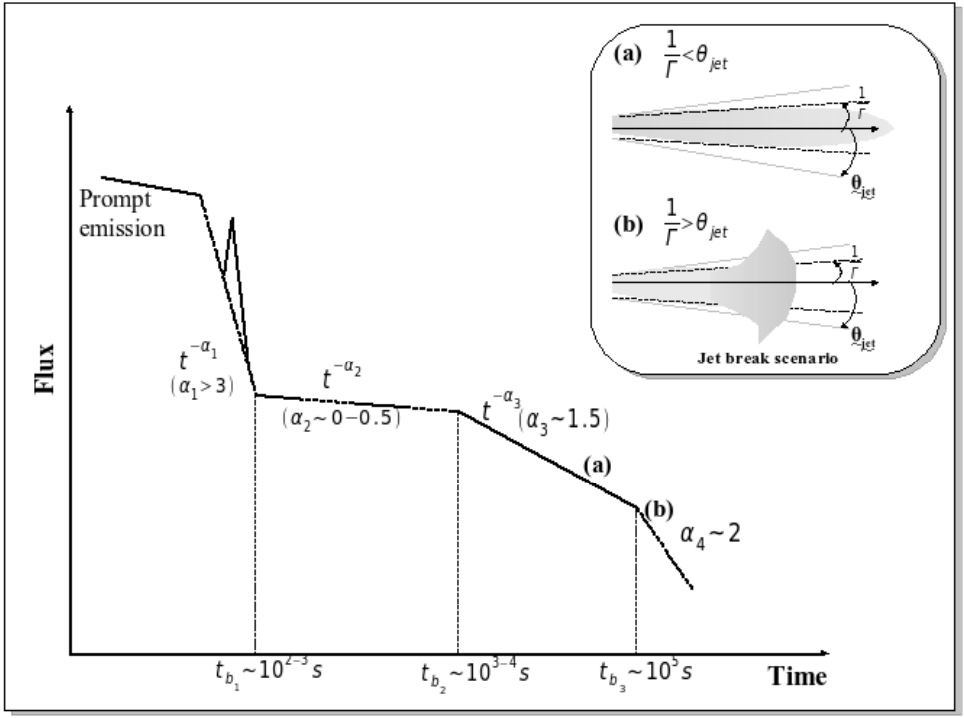}}}

\caption{Scheme showing the X-ray canonical light-curve. 
The dotted segments correspond to temporal sequences that are not
  observed in every GRB light-curve.\label{fig_LC}}
\end{figure*}

From the data collected by the XRT a general pattern rapidly emerged
concerning the shape of the early X-ray afterglow of long GRBs (see
Fig.~\ref{fig_LC}).  Before the standard behavior $F_X\propto t^{-\alpha}$
with $\alpha=1\,-\,1.5$ which was known before {\it Swift}, the canonical
light curve first shows a steep decay (with $\alpha=2\,-\, 5$) followed by a
shallow plateau ($\alpha=0\,-\,1$) that finally breaks to the pre-{\it Swift}
slope after a few hours. Flares with sharp rise and decay times are often
superimposed to this global evolution.

Between the prompt emission and the initial decay phase, a hard to soft
evolution is generally observed (e.g. \cite{OBrien06}). 
Then, a soft to hard jump is often seen at the beginning of the plateau phase \cite{OBrien06}
which indicates the emergence of a new emission
component. The spectral index remains approximately constant over the break
between the plateau and the normal afterglow phase. This break 
therefore does not correspond to a spectral break
(e.g. \cite{Zhang06,Willingale07}). The flares share many common properties
with the pulses of the prompt emission
\cite{Burrows05,Chincarini06,Falcone06,Liang06} (general shape,
brightness-hardness correlation, i.e. on the rise of the flare the spectrum is
harder than during the decay) except that late flares last longer
\cite{Curran09} (with $\Delta t/t\sim 0.1$ -- $0.3$, $\Delta t$ being the
flare duration and $t$ its time of occurence).

\begin{figure*}\center{
\begin{tabular}{cc}
\rotatebox{-90}{\resizebox{.35\hsize}{!}{\includegraphics{XRTLC_GRB060729.ps}}}
& \rotatebox{-90}{\resizebox{.35\hsize}{!}{\includegraphics{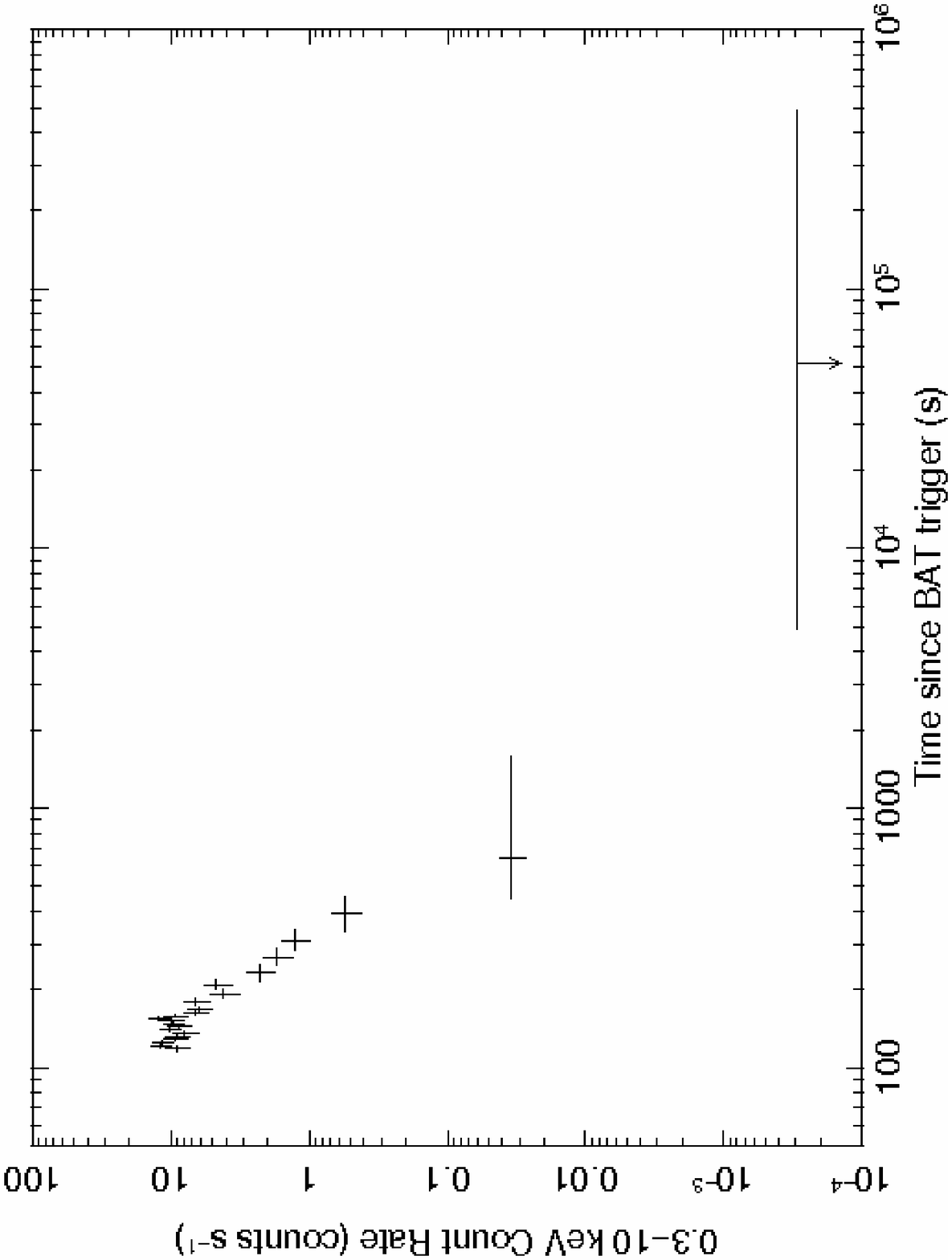}}}\\

\rotatebox{-90}{\resizebox{.35\hsize}{!}{\includegraphics{LC_GRB070110.ps}}}
& \rotatebox{-90}{\resizebox{.35\hsize}{!}{\includegraphics{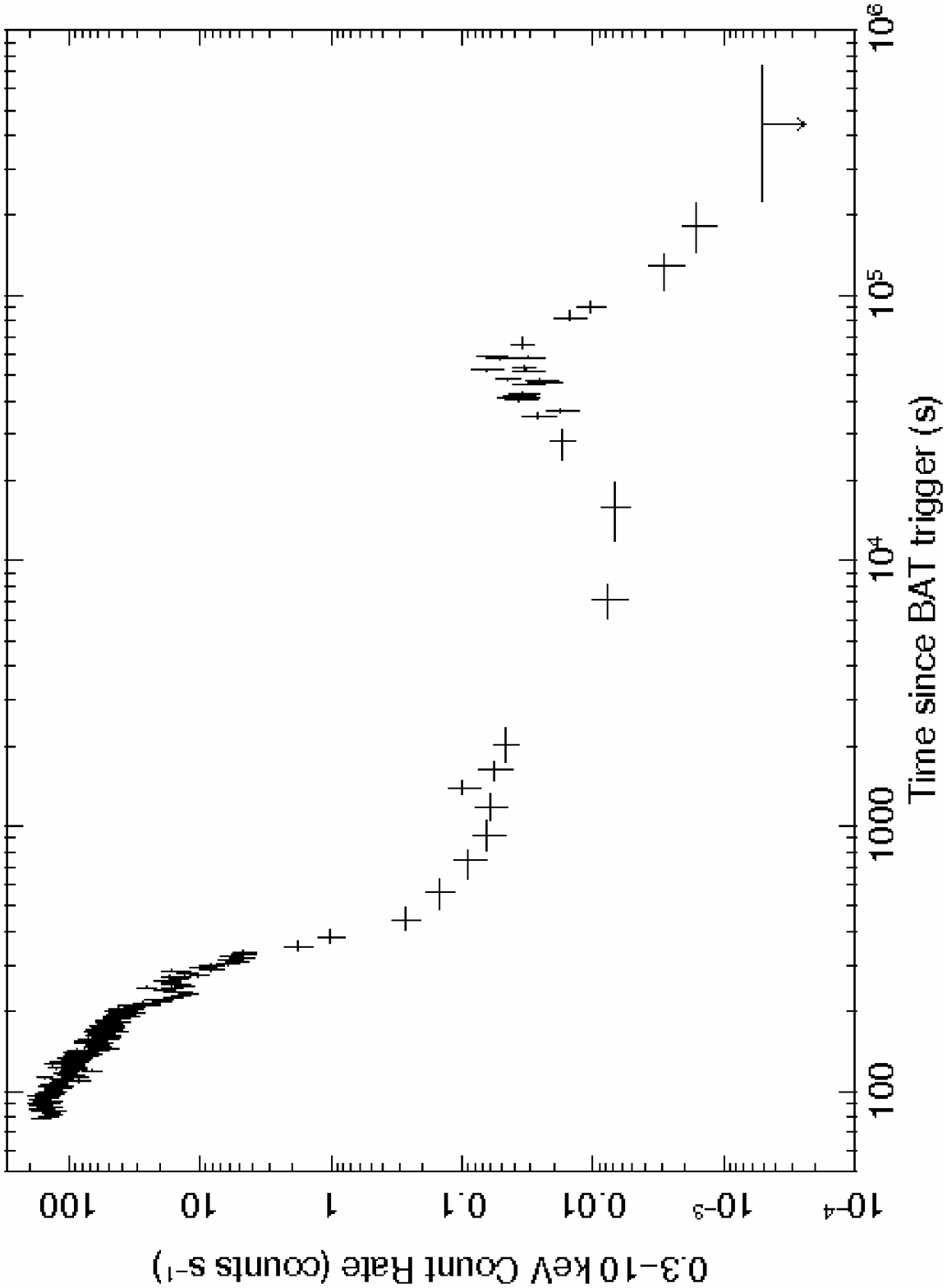}}}\\

\rotatebox{-90}{\resizebox{.35\hsize}{!}{\includegraphics{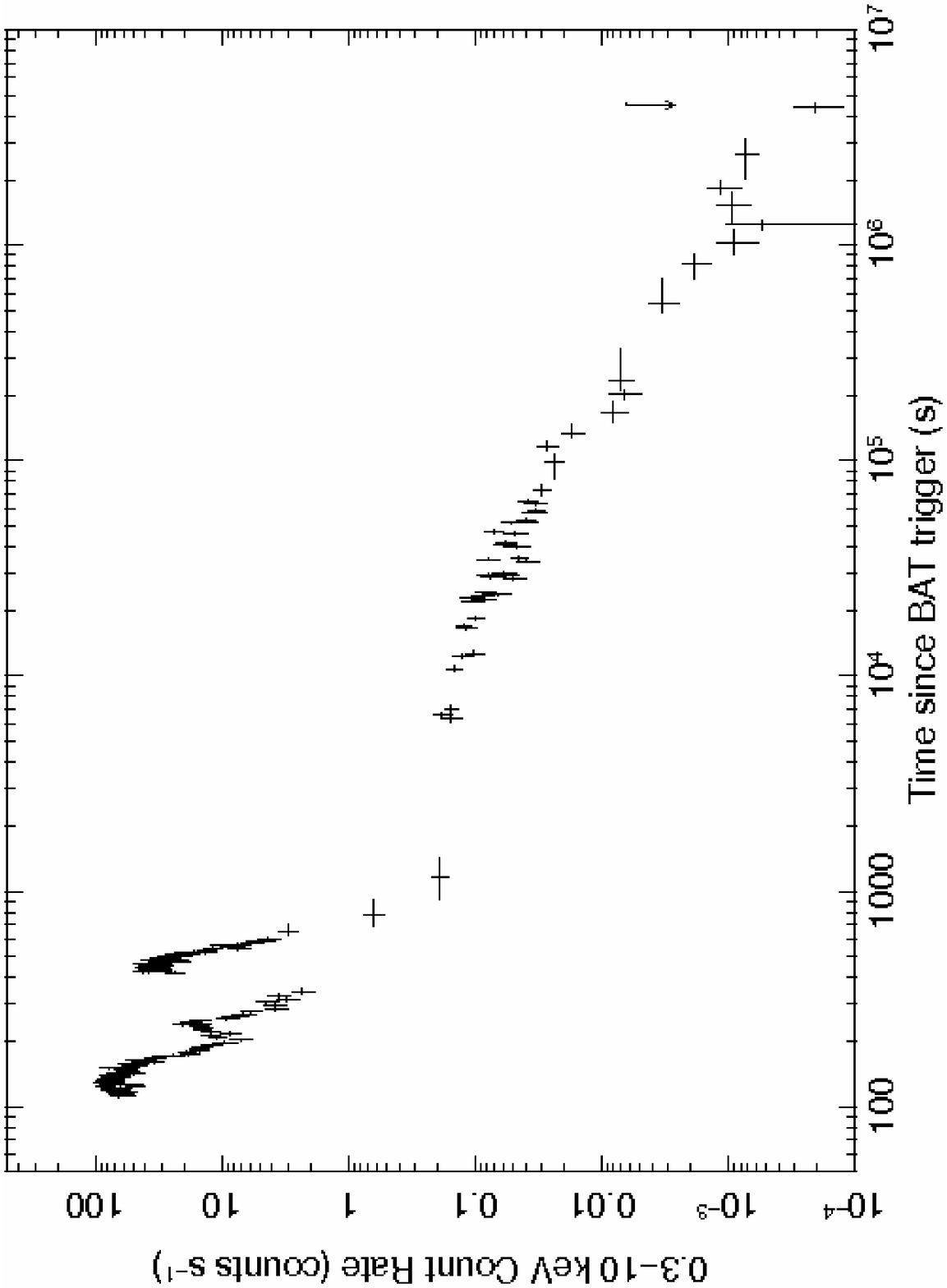}}}
& \rotatebox{-90}{\resizebox{.35\hsize}{!}{\includegraphics{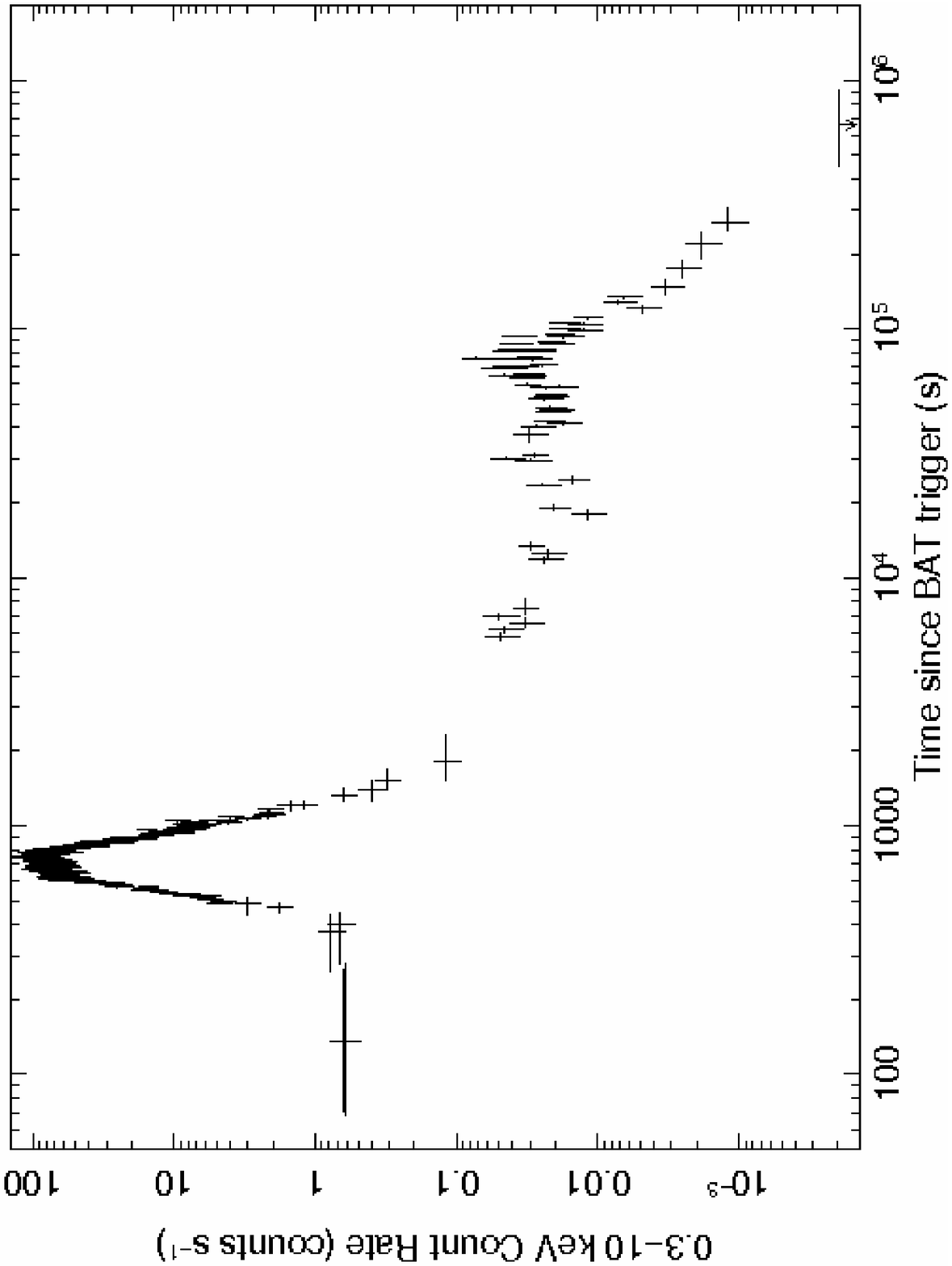}}}\\

\end{tabular}}
\caption{Examples of {\it Swift}-XRT light-curves: GRB 060729 (extended plateau; Top left);
GRB 050421 (no afterglow; Top right); GRB 070110 (flat plateau ended by a steep drop; 
Middle left); GRB 050724 (short burst; Middle right); XRF 050822 (X-Ray Flash \cite{AB11}; Bottom left); 
GRB 050502B (giant flare; Bottom right). The light-curves were obtained using the on-line lightcurve generator
from \cite{Evans09}. \label{fig_example_lc}}
\end{figure*}

The initial steep decay, plateau and flares components are not always present.
Flares are observed in about 50\% of the bursts. The plateau can be absent and
the afterglow then follows a single power law already from the beginning of
the XRT observations (the most extreme case being GRB 061007 \cite{Schady07}
which maintained a constant slope $\alpha=1.6$ from 100 s to more than 10 days
after trigger). Conversely, the plateau can be very flat and extended as in
GRB 060729 \cite{Grupe07} where it lasted nearly one day
(Fig.~\ref{fig_example_lc}).

\begin{figure*}\center{
\begin{tabular}{cc}
\rotatebox{0}{\resizebox{.5\hsize}{!}{\includegraphics{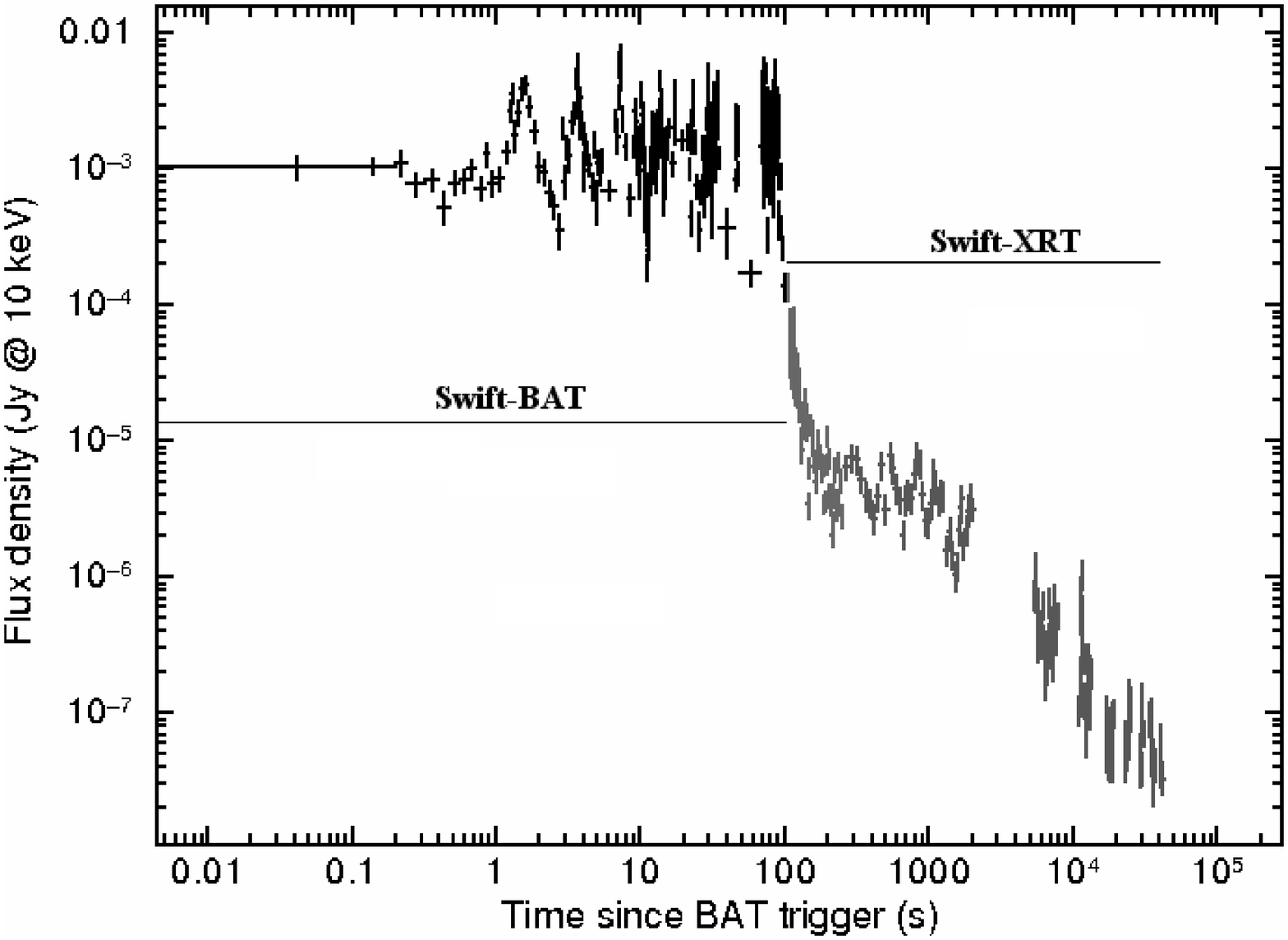}}}
& \rotatebox{0}{\resizebox{.5\hsize}{!}{\includegraphics{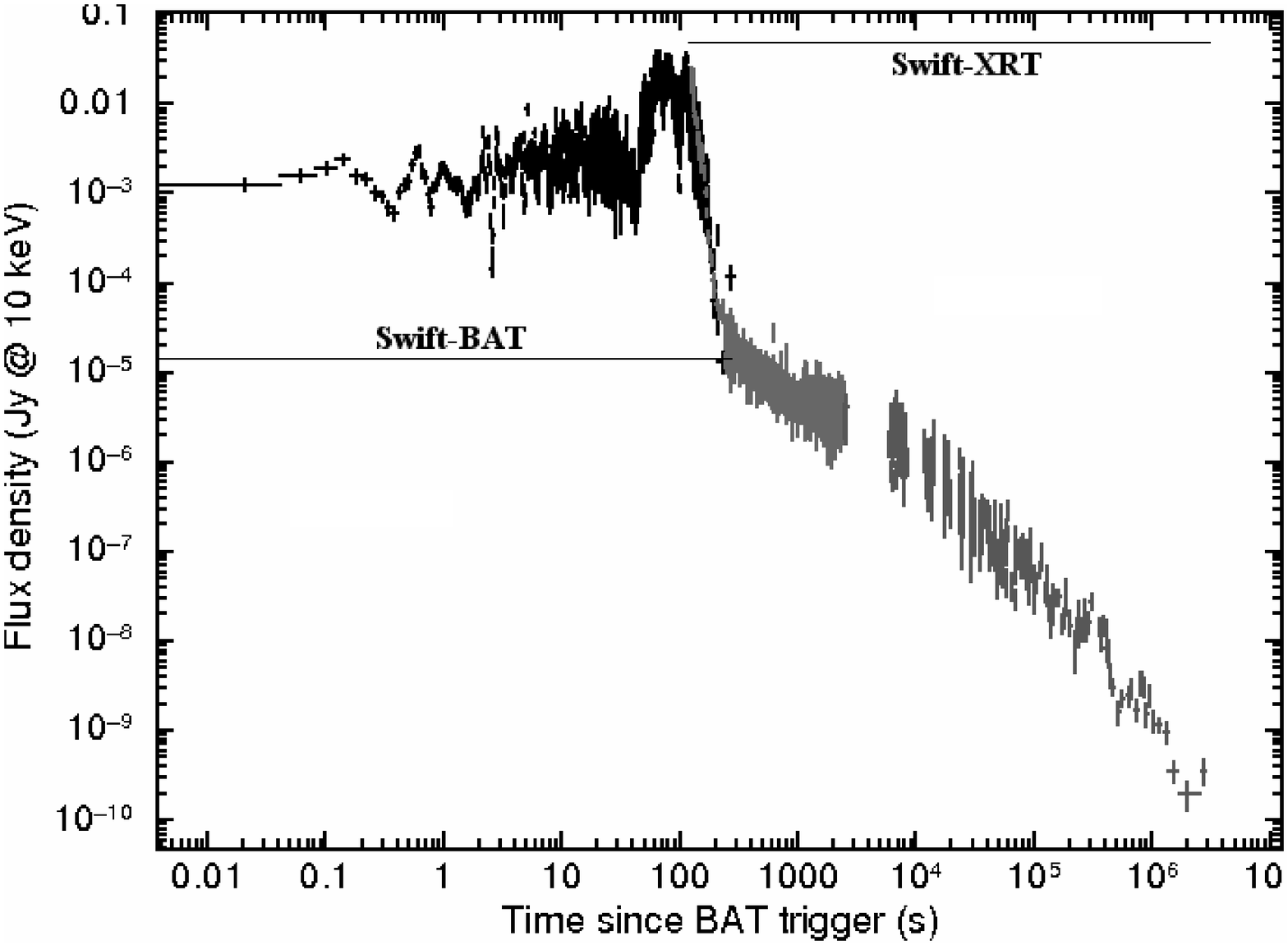}}}\\

\end{tabular}}
\caption{Examples of {\it Swift} combined BAT-XRT lightcurves showing the smooth
connection between the prompt and steep decay of the early afterglow: GRB
080328 (Left); GRB 090618 (Right). The light-curves were obtained using the
on-line lightcurve generator from \cite{Evans09}.\label{fig_example_lc2}}
\end{figure*}

\subsubsection{Implications for afterglow models}
The features of the early X-ray afterglow described above were quite
unexpected and represent difficult challenges for the theoretical models.

\begin{itemize}
\item[$\bullet$] {\it Initial steep decay}: this phase might be the easiest to
explain in the context of the standard model where the prompt emission comes
from internal shocks taking place inside a relativistic outflow emerging from
the central engine.  When internal shocks subside the last shocked shell
flashes as a whole but, due to its curvature, the radiation emitted from
regions away from the line of sight arrives to the observer with a delay, $\Delta t \sim R/2 c\Gamma^2$
where $R$ is the shell radius and $\Gamma$ its Lorentz factor.  
It also benefits less from the ``Doppler boost'',
which amplifies the flux in the direction of fluid motion. Consequently the
flux received by the observer decreases.  It can be shown that it follows a
steep power law decay of the form $F_{\nu}\propto t^{2+\beta}$ where
$\beta\sim 1$ -- 2 is the spectral slope at $\nu$ \cite{Kumar00}. This behavior is in global
agreement with the data
\cite{Zhang06,Tagliaferri05,Goad06,Liang06}. Figure~\ref{fig_example_lc2}
displays two examples of {\it Swift} combined BAT-XRT light-curves that show the
smooth connection between the end of the prompt emission and the steep
decay. 
As can be seen in Fig. 5, slopes steeper than 4 are sometimes observed 
that could be
explained by a shift in the origin of time for the last pulses in the prompt
light curve (i.e. taking the origin of time at the beginning of the pulse and
not at the trigger). 

In some cases, spectral evolution is observed during the
initial steep decay with no corresponding change in the temporal slope even in
XRT light-curves without X-ray flares superimposed
(e.g. \cite{Moretti08}). This spectral evolution may be due to other emission
components coming from shock breakout or the rise of the afterglow, for instance.

Alternatives to internal shocks may explain the initial steep decay of the
afterglow in the same way as long as the radius where the dissipation phase
ends exceeds $10^{15}$ cm. This could be for example the case for some
reconnection models in a magnetized ejecta such as the ``ICMART'' model
\cite{Zhang11}. Conversely if the prompt
emission comes from comptonization (multiple inverse Compton scatterings) 
near the photopheric radius ($10^{11}$ --
$10^{12}$ cm) \cite{RM05,Belo10} the delay $\Delta t$ induced by the curvature of the emitting shell
will be much too short and the prompt phase will stop very abruptly. The
observed initial decay of the afterglow should then result from an effective
behavior of the central engine rather than from geometry. Another alternative
to the curvature emission model invokes the breakout of the jet cocoon from
the stellar envelope \cite{Peer06}.

\item[$\bullet$] {\it Plateau phase}: if the afterglow results from the
forward shock propagating in the circumburst medium the most natural way to
explain the plateau phase is to suppose that energy is added continuously to
the shock so that the decline of the afterglow becomes temporarily shallower
\cite{Zhang06,Nousek06}.  This can be done in two different ways: ({\it i})
the central engine emits in a short time (the duration of the prompt phase)
material with a distribution of Lorentz factors extending down to small
values.  As this ejecta is progressively decelerated the slower material is
able to catch up and adds part of its energy to the forward shock; ({\it ii})
the central engine is active for a longer time (the duration of the plateau
phase corrected for cosmic time dilation) and continuously emitting fast
material that can also catch up and add energy to the shock. 


Energy injection can momentarily reduce the afterglow decline, in agreement
with observations. Two potential difficulties however arise from this
assumption. First, the kinetic energy which is measured from multiwavelength
fits of the afterglow now includes this late supply since the fits are
performed after a few hours. Then, the energy which was available in the
outflow during the prompt phase is reduced, which imposes an increased
efficiency to the prompt mechanism in order to produce the same observed
output in gamma-rays.  In the case of bursts with an extended plateau this
imposes an efficiency $E_{\gamma}/E_{\rm K}>50\%$, out of the reach of most
models invoked to explain the prompt emission. The second potential problem
only concerns the second mode ({\it ii}) of energy injection as it supposes
that the central engine remains active for up to one day to sustain the most
extended plateaus.  This might be possible
\cite{King05,Perna06,Proga06,Rosswog07} but clearly represents a
strong constraint on the central engine, especially for short bursts.
 
Alternatives to energy injection invoke: ({\it i}) a delayed energy transfer
from the forward shock to the electrons from the ISM that gives rise to
a long and slow rise of the afterglow \cite{KobayashiZhang07}; ({\it ii})
an off-beam jet so that when the jet decelerates the visible brightness
surface enters the observer line of sight leading to the rise of the afterglow
\cite{EichlerGranot06}; ({\it iii}) making the afterglow 
with a long-lived reverse shock that would result
from an ejecta having a low Lorentz factor tail (\cite{GDM07,UB07}, see Sect.3.3.1 below).

Some plateaus are even more puzzling as they end with a very steep decay, the best
example being GRB 070110 where $F(t)\propto t^{-9}$ (Fig.4). 
As for flares (see below), this strong variability on a short timescale 
excludes a forward shock origin. It has therefore been proposed that 
these peculiar plateaus result from
some kind of dissipation occurring at small radii, in a persistent 
outflow produced by an active magnetar 
(they are then called ``internal plateaus'' \cite{BZ11,OB09}).  

\item[$\bullet$] {\it Flares}: it appears very unlikely that flares are
produced by the forward shock. Density clumps in the external medium being
swept up by the shock only produce weak bumps with rise and decay times which
are much too long \cite{NG07}. A fast shell
emitted by the central engine and adding its energy to the forward shock
(refreshed shock) would lead to an increase of the underlying afterglow level
after the flare which is not observed \cite{GNP}. As a result, it is generally thought
that flares originate from a late activity of the central engine
\cite{Zhang06,King05,Perna06,Proga06,Rosswog07}. This is supported by the fact
that flares are similar to the pulses observed during the prompt phase except
that late flares last longer, which is not the case for the prompt
pulses. This imposes a very specific behavior to the central engine that will have
to be explained (see \cite{Laz11} for one possibility).
\end{itemize}  

\subsection{The afterglow at lower wavelengths}
\subsubsection{Optical}
\label{optical}

During the pre-{\it Swift} era reasonable joint fits of the X-ray, optical and
radio data have been obtained for many bursts and used to constraint the
density of the external medium and the burst parameters
(e.g. \cite{Panaitescu04a,Panaitescu04b,Panaitescu01}). Due to the limited
sensitivity of the X-ray instruments in BeppoSAX and HETE 2, the optical data
was more complete than the X-rays which generally went below the detection
limit after one day.  The situation changed with the {\it Swift} XRT
instrument which has been able to detect an afterglow in more than 90\% of the
bursts, now leaving many events with uncomplete optical coverage. 

Pre-{\it Swift} optical observations seem to suggest the existence of two
classes of optical afterlows: dark and bright ones \cite{Rol05,Jakobsson04}.
It is worth noting that such a dichotomy is not observed in X-rays.  
Optical observations in the {\it Swift} era have confirmed the existence of these 
dark GRBs \cite{Roming06, Cenko09} (i.e. a
class of GRBs with very faint optical afterglows compared to X-rays even when
observing at early times). Recent works \cite{Melandri08,Perley09,Greiner11}
suggested that a fraction of these dark GRBs may be high-$z$ events while another
fraction could be due to absorption in the vicinity of the GRB progenitor.

The multiwavelength behavior of the early afterglow has revealed a great
complexity: in some bursts the X-ray and optical light curves match together
(following the plateau phase and breaking to an identical slope at the end)
while in others the optical light curve keeps the same power law slope beyond
the end of the plateau in X-rays
(e.g. \cite{Melandri08,PV08,Roming09,Kann10}). These {\it chromatic breaks}
are a challenge to models. In the standard forward shock scenario, they may
require that the microphysical parameters vary with time in a very specific
way (e.g. \cite{Panaitescu06}) or that the X-rays and optical do not come from
the same location or/and radiation process (e.g. \cite{G09}). Models have for
example been proposed where the X-ray emission results from the
up-scattered forward shock radiation \cite{P08}.

A more radical point of view consists to explain the whole afterglow by a
long-lived reverse shock \cite{GDM07,UB07}, which supposes that the source has
produced an ejecta with a low Lorentz factor tail and that the forward shock
is radiatively inefficient. This can be the case if the shock dissipated
energy remains in the baryonic component with almost no transfer to accelerate
electrons or/and amplify the magnetic field in the shocked medium (i.e. very
small $\epsilon_e$ or/and $\epsilon_B$) . If this condition is satisfied the
reverse shock may represent an interesting alternative to the standard
scenario, accounting for many of the observed peculiar temporal and spectral
features.
  
It appears in conclusion that the diversity in the multiwavelength behavior of
the early afterglow makes the modelling of this phase very difficult. It
contrasts with the relative simplicity of the afterglow which was known in the
pre-{\it Swift} era.  The forward shock scenario which seemed able to explain the
data after a few hours does not work (at least in its simplest version) at
earlier times. It is still not clear if it can be adapted with some new
ingredients -- injection of energy, varying microphysics parameters -- or if a
true change of paradigm -- afterglow made by the reverse shock, different
origins for the X-ray and optical emissions -- will be necessary.


\subsubsection{Radio}
Due to self-absorption at early times the radio follow-up of GRBs typically
starts after a few days but can extend to several months and even years (for
instance GRB 030329 \cite{030329}). At these late times it is expected that
the forward shock becomes non-relativistic and that the jet expands laterally
eventually reaching a quasi-spherical shape. In that stage the whole shocked
material is directly accessible to observation and its energy content can be
estimated. This so-called radio calorimetry has yielded typical values for the
total energy release of a few $10^{51}$ erg \cite{FWK00,F05}.

VLBI radio observations have been able to resolve the afterglow of some bursts
and then used to follow the evolution of its apparent size as a function of
time. Comparison to models provide constraints on the energy and opening angle
of the jet as well as on the nature of the external medium (uniform density or
stellar wind). For the few events for which such an analysis has been possible
a (beaming corrected) value of the kinetic energy $E_{\rm K}\sim 10^{51}$ erg
of the outflow has been found, in agreement with the estimate from
calorimetry \cite{ONP04,P07}.

A possible difficulty with the standard afterglow model arises at early times
(1 - 10 days) where the observed radio temporal slopes are often very shallow
and do not agree with (at least the simplest) theoretical predictions 
\cite{PK04}. \cite{PK04} discuss possible solutions to this problem and favor a
scenario where the emission comes from a long-lived reverse shock propagating
in material continuously outflowing from the central source.

\subsection{High energy long lasting emission}

{\it Fermi} LAT (Large Array Telescope \cite{LAT}) observations have shown
that the high energy emission (above 100 MeV) typically starts after the
emission in the keV - MeV range (as seen by the GBM -- Gamma-ray Burst Monitor
\cite{GBM}) and lasts longer. It is not clear if the data correspond to the
afterglow already from the beginning or if it should be explained by a
superposition of prompt and afterglow components.  Some hints of variability
in the GeV range in possible coincidence with the keV -- MeV light curve may
indicate a prompt origin for the early GeV emission \cite{Z11} but then one
has to explain why the subsequent afterglow component connect almost perfectly
with the prompt phase. Indeed, in the few bursts for which LAT data cover both
the prompt phase and early afterglow, no steep decay is observed at the end of
the prompt phase (except maybe in GRB 090510 \cite{GRB090510,GGN10}) in contrast
with what is seen in X-rays.

Moreover the radiative process at the origin of the high energy emission is
not known. Some spectra can be well fitted by a broken power law (Band
function \cite{Band93}) from the keV to GeV range while others require an
additional power law or a thermal component.  Most models invoke synchrotron
or/and Inverse Compton emission of shock accelerated electrons to explain the
data, but hadronic models where the GeV photons and the additional power law
come from hadronic processes have also been considered \cite{AGM09}. Only one
burst (the short GRB 090510) has been simultaneously seen by {\it Swift} and
the {\it Fermi} LAT starting from the trigger time. More such common
detections can be expected in the future and should help to better constrain
the radiation processes at work.

More details on the high energy emission of GRBs will be found in \cite{PC11}
(this issue).

\section {The afterglow of short GRBs and X-ray flashes}
\subsection{The afterglow of short GRBs}

Short bursts tend to be harder than long bursts (e.g. \cite{Preece00}) and
emit less photons which made their detection and localization more difficult.
The first afterglows of short bursts were found only in 2005 (one by HETE-2 --
GRB 050709 \cite{GRB050709} -- and all the others by {\it Swift}
\cite{Gehrels05,Barthelmy05,Zhang07} and references therein).

The nature of the host galaxies, the location of the GRBs within the host
galaxies and the absence of a supernova imprint in the visible light curve
(even when the host is located at a redshift below 0.5) suggests that the
progenitors of short bursts are different from those of long bursts.  Several
short bursts are clearly associated to elliptical galaxies while others with
accurate positions appear to have no coincident host, which clearly excludes
progenitors belonging to the young population and favors merger scenarios
involving compact objects -- two neutron stars or a neutron star-black hole
binary \cite{Paczynski86,Eichler89,Rosswog03}. Short GRBs are located at
lower redshift ($\langle z\rangle \sim 0.4$) than long ones
but recently, a population of high-$z$ ($z > 1$) short GRBs may have 
been detected (see \cite{Berger07}).

Afterglows of short bursts fall in two classes: about 40\% of the events have
no detectable afterglow after about 1000 s while the other 60\% have long
lasting afterglows comparable to those of long GRBs. Short GRBs appear as less energetic
events when compared to long GRBs with a gamma-ray jet corrected energy around
$10^{49}$ erg (against $10^{51}$ erg for long GRBs (e.g. \cite{Burrows06} for
instance). If indeed short bursts result from the merging of two compact
objects the kick received when the neutron star or black hole components
formed in supernova explosions allows the system to reach the low density
outskirts of the host galaxy (or even to leave the galaxy) before coalescence
occurs. This can naturally explain why a fraction of the afterglows are so dim
or have no coincident host. It is interesting to note that in the case of a
few long GRBs (such as GRB 050421 \cite{Godet06,Vetere09} - see Fig.4) 
no afterglow or a dim X-ray afterglow was detected even if
{\it Swift} started promptly observing the GRB position after the {\it
Swift}-BAT trigger, which might suggest that these events are ``naked GRBs''
with ISM density as low as that observed for some short GRBs
\cite{Zhang07}.

Some flaring activity has also been seen in short bursts up to several
hundreds seconds after trigger such as in GRB 061210 or GRB 100117A. In
addition to flares a wide bump was also present at much later times (between
$5\times 10^4$ and $10^5$ s) in GRB 050724 (see Fig.4). If flares result from a persistent
activity of the central engine this would imply that it can still operate even
in bursts where the prompt phase lasted less than one second (for instance 200
ms in GRB 061210 \cite{cumin06}).

The multi-wavelength observations of afterglows of short GRBs also led to
constrain the opening angle of the jets that appear to be less
collimated than in long GRBs \cite{Burrows06,Watson04}.

\subsection{X-ray flashes}

X-ray flashes (XRFs) share many common properties with GRBs except
that their spectrum peaks in the X-ray range (from a few keV to a few tens of
keV). Observations of XRF afterglows have helped to answer several questions
regarding their nature and origin. First, they confirmed that XRFs are, as
GRBs, located at cosmological distances. The first redshift was measured for
XRF 020903 at $z=0.25$ \cite{Soderberg04} but larger values (up to $z=2.65$
for XRF 030429 \cite{J04}) were later obtained.

Another issue was to decide if XRFs were intrinsically soft or if their low
$E_{\rm p}$ (the peak energy of the $\nu\,F_{\nu}$ spectrum) resulted from
special observational circumstances.  A first possibility, a large redshift,
was clearly excluded already with XRF 020903 which had an $E_{\rm p}$ of a few
keV at $z=0.25$. It was also suggested that XRFs were soft events because they
were seen off-axis with a reduced Lorentz boost in the direction of the
observer.  Such a geometry delays the rise of the afterglow until the angle
$\theta_j+1/\Gamma$, where $\theta_j$ is the physical opening angle of the jet
and $1/\Gamma$ the relativistic beaming angle (which is increasing 
as a result of
deceleration) can encompass the line of sight. Afterglow observations can then
be used to test the off-axis scenario but the available data has not yet
provided a fully clear conclusion.  XRF 030723 \cite{Butler05} and XRF 080330
\cite{Guidorzi09} seem to be consistent with an off-axis jet while XRF 020903,
050215B \cite{Levan06}, 050406 \cite{Romano06}, 050416A \cite{Mangano07} (see
also \cite{Holland07}) and 050822 \cite{Godet07} appear to have been
intrinsically soft. \cite{Lamb05} also suggested that the softness of XRFs is
due to a wider jet opening angle compared to classical
GRBs. \cite{Mangano07,Godet07} showed in the case of XRF 050822 and 050416A
that the jet opening angle is likely to be more than $20^\circ$, which is much
larger than the average value $\theta_j\sim 5^\circ$ derived for GRBs by
\cite{Frail01}.

\section{Current issues and problems}
In addition to the questions raised by the early afterglow or the high energy
emission we list below a few additional problems faced by standard afterglow
modelling:

\subsection{The jet break problem}
\label{jet}
As the ejecta is decelerated by the external medium relativistic beaming of
the emitted radiation becomes less effective and $1/\Gamma$ becomes eventually
larger than the physical opening angle of the jet $\theta_j$
\cite{Rhoads99,Rhoads97}. This leads to an achromatic break in the light curve
(since it is a purely geometrical effect -- see Fig.~\ref{fig_LC}). The time
$t_j$ of this ``jet break'' can be used to estimate $\theta_j$ and therefore
the true energy release of the burst, corrected for beaming
\cite{Frail01,LZ05}. In the pre-{\it Swift} era, candidate jet breaks were
identified at typical times $t_j\sim$ one day and yielded values of the energy
in gamma-rays $E_{\gamma}$ of a few $10^{50}$ erg, suggesting the existence of
a standard energy reservoir \cite{Frail01}.

The situation became less clear in the {\it Swift} era as it is now possible
to follow some X-ray afterglows for several weeks and even months.  In many
cases no break is observed (e.g. \cite{Willingale07,Liang09}, see however
\cite{Racusin09}); which seems to indicate beaming angles and emitted energies
larger than previously expected.  In a study of 5 bright bursts \cite{Cenko10}
were able to identify late jet breaks at about 10 days. The corresponding
beaming angles were all larger than 5$^{\circ}$ (with $\theta_j > 20^{\circ}$
in one case). The true energy in gamma-rays then reaches $10{^{52}}$ erg for
these bright bursts, much above what was believed to be the standard value.
One possible way to avoid increasing too much the energy budget of the central
engine would be to suppose that the radiation is not isotropic in the comoving
frame of the emitting material \cite{BDMU10}. Then, it would be beamed in an
angle smaller than $1/\Gamma$ in the observer frame and, for a given physical
opening angle of the jet, the jet break will be delayed.

It should finally be mentioned that the new situation revealed by {\it Swift}
casts doubts about the early jet breaks from the pre-{\it Swift} era.  It
cannot be excluded that at least some of them corresponded to the end of the
plateau phase and were misinterpreted as ``jet breaks''.

\subsection{Nature of the circum-burst medium}

It has been mentioned in Section~\ref{model} that, in the framework of the
standard model, multi-wavelength fits of afterglows seem to favor a low density, uniform
external medium, even for long bursts (e.g. \cite{Zhang06,PK01,Schulze11}). This is
somehow unexpected since long bursts are supposed to be produced by some
extreme kind of type Ib/c supernovae \cite{Woosley93}. The progenitors of
type Ib/c are Wolf-Rayet stars which normally have strong winds. 
One possible solution could be that in most cases the radius of the
wind-termination shock is relatively small so that the jet
deceleration mostly happens after this radius.  
The requirement of the collapsar model that the 
progenitor star keeps a large angular momentum up to its collapse indeed 
implies a lower mass loss \cite{Langer10}, which is consistent with the
possibility of a small termination shock radius.

It nevertheless
remains surprising that more weight is often given in the literature to the
results of a modelling process that rely on many uncertain parameters and
assumptions (such as taking constant values of $\epsilon_e$ and $\epsilon_B$)
than to standard astronomical knowledge...

\subsection{Microphysics}

We just indicated that most of the modelling work is performed by adopting fixed
values of the microphysics parameters.  This is the most natural zeroth-order
assumption but it is not necessarily true. The problem is that, unfortunately,
shock physics has not reached the status where reliable predictions can be
made on how these parameters should vary with the shock strength or density of
the external medium. Allowing $\epsilon_e$ and $\epsilon_B$ to vary obviously
adds considerable freedom to the modelling process but also makes it very
uncertain with limited predictive power. For example the often used ``closure
relations'' \cite{Zhang06} that link the temporal and spectral indices in the different
regimes of afterglow evolution rely on the assumption that the microphysics
parameters stay constant. The preference for a uniform external medium also
results from the same assumption. If the microphysics parameters can vary,
wind models can be more easily accomodated.

Also, the distribution of the power-law index $p$ of the electron
population appears far from universal \cite{Zhang06,Shen06},
contrary to what is expected from theoretical calculations and
simulations of particle acceleration in relativistic shocks
($p=2.2-2.3$). This is puzzling but could also be a consequence of 
inappropriate assumptions regarding $\epsilon_e$ and $\epsilon_B$.

\subsection{Consistency of the global model}
It should finally be reminded that any new ingredients added to one part of
the model should be consistent with the rest of it! In that respect some of
the solutions proposed to solve problems in afterglow physics have important
consequences for the prompt phase (and vice-versa).  \\

\begin{itemize}

\item Initial steep decay: it is tempting to interpret the initial steep decay
at the end of the prompt phase as the high latitude emission of the last
flaring shell. This supposes that the delay $\Delta t=R/2 c \Gamma^2$ ($R$ being 
the radius of the shell and $\Gamma$ its Lorentz factor)
is not much smaller that the duration of the prompt phase. This condition is
naturally satisfied by internal shocks but not by a dissipative
photosphere and by most magnetic reconnection models where $\Delta t$ is too
short.

\item Energy injection is generally invoked to explain the plateau phase of
the early X-ray afterglow. We already explained why such a scenario implies a
very high efficiency for the prompt phase which practically excludes internal
shocks and challenges most other models. Moreover the propagation of the
reverse shock through the material adding its energy to the forward shock will
add a new contribution to the early afterglow which has never been properly
estimated.

\item Models where the afterglow is made by the reverse shock obviously imply
that a reverse shock is present which is not the case in a magnetized ejecta
where the magnetization $\sigma = E_{\rm mag}/E_{\rm K} > 1$. Therefore such
models either require that the outflow emerging from the central engine is not
magnetized or that most of the magnetic energy has already been converted to
kinetic energy when the reverse shock develops.

\end{itemize}

\section{Conclusion}

Following the discovery of the afterglows by Beppo-SAX, it was generally
believed that their physics would be much simpler than that of the prompt
phase of GRBs. The basic framework involved a relativistic generalization of
the classical Sedov problem and part of the theory (the Blandford-McKee
solution) even preexisted the 1997 discovery. The situation became more
confuse in the {\it Swift} era when the early afterglow was first observed.  The
standard forward shock model (at least in its basic version) appeared unable to
account for the new phenomenology (the plateau phase and the flares) revealed
by {\it Swift}.

Solutions supposing a long term activity of the central engine that have been
proposed to explain both the plateau and the flares are not free of pitfalls:
they increase the required efficiency of the prompt phase to levels that might
be incompatible with most models, especially internal shocks. Flares impose a
specific, intermittent behavior to the source that is different from the more
continuous and regular activity needed to sustain the plateau.

Proposed alternatives to the standard model rely on new assumptions that will
have to be confirmed by further confrontation to the data.  For instance, the
role played by the reverse shock remains uncertain. The reverse shock may
simply not exist if the flow is still highly magnetized when it starts to be
decelerated. It will be short lived if the ejecta only contains material of high
Lorentz factor and then might be responsible for the early optical flashes that
have been seen in a few bursts. Finally if the ejecta has
a low Lorentz factor tail, the reverse shock will be long-lived and may
possibly contribute to the afterglow.

{\it Fermi} LAT observations also are not easy to interpret. The high energy
emission extends beyond the end of the prompt phase and should somehow be
linked to the afterglow, at least at late times. However it is not clear if it
simply corresponds to the high energy tail of the usual afterglow or if it
involves a new, possibly hadronic component.

It finally appears that the results from the {\it Swift} and {\it Fermi}
satellites have not contributed to clarify our understanding of afterglow
physics. Instead they have added new puzzles and raised new questions!  This
may appear as a frustrating and pessimistic conclusion but one should bear in 
mind the difference between progress in understanding and progress in
knowledge.  In the past fifteen years we have made tremendous progress in our
knowledge of GRBs. This is especially true for afterglows which simply ``did
not exist'' before 1997. We have now collected a huge amount of
multiwavelength data starting from one minute after the trigger.
Future
missions like SVOM (to be launched in 2016) will continue this quest.  
SVOM will not only obtain GRB key parameters such as $E_{\rm p}$ 
and the redshift, but will also provide
well-sampled multi-wavelength light-curves from early to late times. In
addition to a fast slew capability of the satellite and sensitive
multi-wavelength NFIs providing accurate positions, this
imposes to build a strong synergy between space and ground via an
international network of robotic telescopes. This will ensure that large
facilities such as the VLT are able to perform observations as quickly
as possible. 

One can then hope that increasing our knowledge of GRBs will ultimately lead to more
understanding...





\end{document}